%% file: papier.tex
\documentclass{article}
\usepackage{spconf}
\usepackage{amsmath}
\usepackage{amsthm}
\usepackage{amsfonts}
\usepackage{amssymb}
\usepackage{graphicx,enumerate}

\include{prelim}

\newcommand{\eps}{\varepsilon}
\newcommand{\Hm}{\mathcal{H}}
\newcommand{\Tr}{^\mathrm{T}}

\renewcommand{\ge}{\geqslant}
\renewcommand{\geq}{\geqslant}
\renewcommand{\leq}{\leqslant}
\newcommand{\reg}{f_2}
\newcommand{\Prj}{\mathcal{P}}

\DeclareMathOperator{\prox}{prox}
\DeclareMathOperator{\rprox}{rprox}
\DeclareMathOperator*{\argmin}{arg\ min}

\graphicspath{{images/}{../../Material/Results/isbi2008/}}

\title{Deconvolution of confocal microscopy images using proximal iteration
  and sparse representations}
\name{F.-X. Dup\'e$^\text{a}$, M.J. Fadili$^\text{a}$ and J.-L. Starck$^\text{b}$}
\address{\begin{tabular}{cc}
        $^\text{a}$ GREYC UMR CNRS 6072 & $^\text{b}$ DAPNIA/SEDI-SAP CEA-Saclay\\
        14050 Caen France & 91191 Gif-sur-Yvette France
        \end{tabular}
}

\begin{document}
%
\maketitle
\begin{abstract}
  We propose a deconvolution algorithm for images blurred and degraded by a Poisson
  noise. The algorithm uses a fast proximal backward-forward splitting iteration. This
  iteration minimizes an energy which combines a \textit{non-linear} data fidelity term,
  adapted to Poisson noise, and a non-smooth sparsity-promoting regularization (e.g
  $\ell_1$-norm) over the image representation coefficients in some dictionary of
  transforms (e.g. wavelets, curvelets). Our results on simulated microscopy images of
  neurons and cells are confronted to some state-of-the-art algorithms. They show that our
  approach is very competitive, and as expected, the importance of the non-linearity due
  to Poisson noise is more salient at low and medium intensities.  Finally an experiment
  on real fluorescent confocal microscopy data is reported.
\end{abstract}

\vspace*{-0.4cm}
\section{Introduction}
\label{sec:intro}
\vspace*{-0.4cm}

Fluorescent microscopy suffers from two main sources of degradation: the optical system
and the acquisition noise. The optical system has a finite resolution introducing a blur
in the observation. This degradation is modeled as convolution with the point spread
function (PSF).  The second source of image degradation is due to Poisson count process
(shot noise). In presence of Poisson noise, several deconvolution algorithms have been
proposed such as the well-known Richardson-Lucy (RL) algorithm or Tikhonov-Miller inverse
filter, to name a few. RL is extensively used for its good adaptation to Poisson noise,
but it tends to amplify noise after a few iterations.  Regularization can be introduced in
order to avoid this issue. In biological imaging deconvolution, many kinds of
regularization have been suggested: total variation with RL \cite{Dey2004} which gives
staircase artifacts, Tikhonov with RL (see \cite{Sarder2006} for a review), etc.  Wavelets
have also been used as a regularization scheme when deconvolving biomedical images;
\cite{Monvel2001} presents a version of RL combined with wavelets denoising, and
\cite{Vonesch2007} use the thresholded Landweber iteration of \cite{Daubechies2004}. The
latter approach implicitly assumes that the contaminating noise is Gaussian.


In the context of deconvolution with Gaussian white noise, sparsity-promoting
regularization over orthogonal wavelet coefficients has been recently proposed
\cite{Daubechies2004,Combettes2005}. Generalization to frames was proposed in
\cite{Teschke2007,Combettes2007b}. In \cite{Fadili2006a}, the authors presented an image
deconvolution algorithm by iterative thresholding in an overcomplete dictionary of
transforms. However, all sparsity-based approaches published so far have mainly focused on
Gaussian noise.

In this paper, we propose an image deconvolution algorithm for data blurred and
contaminated by Poisson noise. The Poisson noise is handled properly by using the Anscombe
variance stabilizing transform (VST), leading to a {\it non-linear} degradation
equation with additive Gaussian noise, see \eqref{eq:3}. The deconvolution problem is then
formulated as the minimization of a convex functional with a non-linear data-fidelity term
reflecting the noise properties, and a non-smooth sparsity-promoting penalty over
representation coefficients of the image to restore, e.g. wavelet or curvelet
coefficients.  Inspired by the work in \cite{Combettes2005}, a fast proximal iterative
algorithm is proposed to solve the minimization problem. Experimental results are carried
out to compare our approach on a set of simulated and real confocal microscopy images, and
show the striking benefits gained from taking into account the Poisson nature of the noise
and the morphological structures involved in the image.

\vspace*{-0.4cm}
\subsection*{Notation}
\label{sec:notation}
\vspace*{-0.4cm}

{\footnotesize
Let $\mathcal{H}$ a real Hilbert space, here a finite dimensional vector subspace of
$\R^n$. We denote by $\norm{.}_2$ the norm associated with the inner product in
$\mathcal{H}$, and $\I$ is the identity operator on $\mathcal{H}$. $\vx$ and $\va$ are
respectively reordered vectors of image samples and transform coefficients. A function $f$
is coercive, if $\lim_{\norm{\vx}_2 \rightarrow +\infty}f\parenth{\vx}=+\infty$.
$\Gamma_0(\mathcal{H})$ is the class of all proper lower semi-continuous convex functions
from $\mathcal{H}$ to $]-\infty,+\infty]$.
}

\vspace*{-0.4cm}
\section{Problem statement}
\label{sec:problem-statement}
\vspace*{-0.4cm}

Consider the image formation model where an input image $\vx$ is blurred by a point spread
function (PSF) $h$ and contaminated by Poisson noise. The observed image is then a
discrete collection of counts $y=(y_i)_{1 \leq i \leq n}$ where $n$ is the number of
pixels. Each count $y_i$ is a realization of an independent Poisson random variable with
a mean $(h \circledast x)_i$, where $\circledast$ is the circular convolution
operator. Formally, this writes $y_i \sim \mathcal{P}\parenth{(h \circledast x)_i}$.

A naive solution to this deconvolution problem would be to apply traditional approaches
designed for Gaussian noise. But this would be awkward as (i) the noise tends to Gaussian
only for large mean $(h \circledast x)_i$ (central limit theorem), and (ii) the noise
variance depends on the mean anyway. A more adapted way would be to adopt a bayesian
framework with an appropriate anti-log-likelihood score reflecting the Poisson statistics of
the noise. Unfortunately, doing so, we would end-up with a functional which does not
satisfy some key properties (the Lipschitzian property stated after (\ref{eq:6})), hence
preventing us from using the backward-forward splitting proximal algorithm to solve the
optimization problem. To circumvent this difficulty, we propose to handle the noise
statistical properties by using the Anscombe VST defined as
\begin{equation}
  \label{eq:2}
  z_i = 2\sqrt{y_i + \tfrac{3}{8}}, ~ 1 \leq i \leq n .
\end{equation}
Some previous authors \cite{ChauxSPIE} have already suggested to use the Anscombe VST, and
then deconvolve with wavelet-domain regularization as if the stabilized observation $z_i$
were linearly degraded by $h$ and contaminated by additive Gaussian noise. But this is not
valid as standard asymptotic results of the Anscombe VST state that
\begin{equation}
  \label{eq:3}
  z_i = 2\sqrt{(h\circledast x)_i+\tfrac{3}{8}} + \eps,\quad \eps \sim \mathcal{N}(0,1)
\end{equation}
where $\eps$ is an additive white Gaussian noise of unit variance\footnote{Rigorously 
speaking, the equation is to be understood in an asymptotic sense.}. In words, $z$ is
{\em non-linearly} related to $x$. In Section~\ref{sec:sparse-iter-deconv}, we provide an
elegant optimization problem and a fixed point algorithm taking into account such a
non-linearity.

\vspace*{-0.4cm}
\section{Sparse image representation}
\label{sec:sparse-image-repr}
\vspace*{-0.2cm}

Let $x \in \Hm$ be an $\sqrt{n}\times\sqrt{n}$ image. $x$ can be written as the
superposition of elementary atoms $\varphi_\gamma$ parametrized by $\gamma \in
\mathcal{I}$ such that $x = \sum_{\gamma \in \mathcal{I}} \alpha_\gamma \varphi_\gamma =
\Phi \va,\quad \abs{\mathcal{I}} = L, ~ L\ge n$.  We denote by $\Phi$ the dictionary
i.e. the $n\times L$ matrix whose columns are the generating waveforms
$\parenth{\varphi_\gamma}_{\gamma \in \mathcal{I}}$ all normalized to a unit
$\ell_2$-norm.  The forward transform is then defined by a non-necessarily square matrix
$\T = \Phi\Tr \in \mathbb{R}^{L\times n}$. In the rest of the paper, $\Phi$ will be an
orthobasis or a tight frame with constant $A$.

\vspace*{-0.4cm}
\section{Sparse Iterative Deconvolution}
\label{sec:sparse-iter-deconv}
\vspace*{-0.2cm}

\subsection{Optimization problem}
\label{sec:sparse-optim}
\vspace*{-0.2cm}

The class of minimization problems we are interested in can be stated in the general form
\cite{Combettes2005}:
\begin{equation}
  \label{eq:6}
  \argmin_{x \in \Hm} f_1(x) + f_2(x) .
\end{equation}
where $f_1 \in \Gamma_0(\mathcal{H})$, $f_2\in\Gamma_0(\mathcal{H})$
and $f_1$ is differentiable with $\kappa$-Lipschitz
gradient. We denote by $\M$ the set of solutions.

\noindent
From \eqref{eq:3}, we immediately deduce the data fidelity term
\begin{gather}
\label{eq:7}
  F \circ H \circ \Phi~(\va), ~ \text{with} \\
  F : \eta \mapsto \sum_{i=1}^n f(\eta_i),\quad f(\eta_i) = \frac{1}{2}
  \left( z_i - 2\sqrt{\eta_i+\tfrac{3}{8}} \right)^2 , \nonumber 
\end{gather}
where $H$ denotes the convolution operator. From a statistical
perspective, \eqref{eq:7} corresponds to the anti-log-likelihood score.

Adopting a bayesian framework and using a standard maximum a posteriori (MAP) rule, our
goal is to minimize the following functional with respect to the representation
coefficients $\alpha$:
\begin{gather}
  \label{eq:9}
  (\P_{\lambda,\psi}): \argmin_{\alpha} J(\alpha) \\
  J : \alpha \mapsto \underbrace{F \circ H \circ \Phi\ (\alpha)}_{f_1(\va)} + 
  \underbrace{\imath_{\C} \circ \Phi\ (\alpha) + \lambda \sum_{i=1}^L \psi(\alpha_i)}_{f_2(\va)}, \nonumber 
\end{gather}
where we implicitly assumed that $(\alpha_i)_{1 \leq i \leq L}$ are independent and
identically distributed. The penalty function $\psi$ is chosen to enforce sparsity,
$\lambda > 0$ is a regularization parameter and $\imath_{\C}$ is the indicator function of
a convex set $\C$. In our case, $\C$ is the positive orthant. We remind that the
positivity constraint is because we are fitting Poisson intensities, which are positive by
nature.

\vspace*{-0.4cm}
\subsection{Proximal iteration}
\label{sec:proximal-iteration}
\vspace*{-0.2cm}

 We now present our main proximal iterative algorithm to solve
the minimization problem $(\P_{\lambda,\psi})$:
\begin{theorem}
  \label{th:2}
  $(\P_{\lambda,\psi})$ has at least one solution ($\M \ne \emptyset$). The solution is
  unique if $\psi$ is strictly convex or if $\Phi$ is a orthobasis and $\mathrm{Ker}(H) =
  \emptyset$.  For $t \geq 0$, let $(\mu_{t} )_{t}$ be such
  that $0 < \inf_{t} \mu_{t} \leq \sup_{t} \mu_{t}< \parenth{\frac{3}{2}}^{3/2}/\parenth{2
    A \norm{H}_2^2 \norm{z}_{\infty}}$.  Fix $\va_{0}\in\C\circ\Phi$, for every $t \ge 0$, set
  \begin{equation}
    \label{eq:5}
    \va_{t+1} = \prox_{\mu_{t}f_{2}}\parenth{\va_{t} -
      \mu_{t}\nabla f_{1}(\va_{t})} ~ ,
  \end{equation}
  where $\nabla f_{1}$ is the gradient of $f_1$
  and $\prox_{\mu_t f_2}$ is computed using the following iteration:
  let $\sum_t \nu_t(1-\nu_t)=+\infty$, take $\gamma^{0} \in \Hm$, and
  define the sequence of iterates:
  \begin{eqnarray}
      \label{eq:proxtframe1}
      \gamma^{t+1} = \gamma^t + \nu_t\parenth{\rprox_{\mu_{t}\lambda\Psi + \tfrac{1}{2}\norm{. - \va}^2}\circ\rprox_{\imath_{\C'}} - \I}(\gamma^t) ,
    \end{eqnarray}
    where $\prox_{\mu_{t}\lambda\Psi + \tfrac{1}{2}\norm{. - \va}^2}
    (\gamma^t)= \parenth{\prox_{\mu_{t}\tfrac{\lambda}{2}\psi} ((\va_i + \gamma^t_i)/2)}_{1\leq i
      \leq L}$, $\Prj_{\C'} = \prox_{\imath_{\C'}} = A^{-1} \Phi\Tr\circ\Prj_\C\circ\Phi
    +\parenth{\I - A^{-1} \Phi\Tr\circ\Phi}$, $\rprox_{\varphi} = 2\prox_{\varphi} - \I$
    and $\Prj_\C$ is the projector onto the positive orthant $(\Prj_\C \eta)_i =
    \max(\eta_i, 0)$. Then,
    \begin{equation}
      \label{eq:proxtframe2}
      \gamma^t \rightharpoonup \gamma ~ \text{and} ~ \prox_{\mu_{t}\reg}(\va) = \Prj_{\C'}(\gamma).
    \end{equation}
    \noindent
    Then $(\alpha_t)_{t \geq 0}$ converges (weakly) to a solution of $(\P_{\lambda,\psi})$.
\end{theorem}
A proof can be found in \cite{Dupe2008}. $\prox_{\delta\psi}$ is given by,
\begin{theorem}
\label{th:3}
Suppose that (i) $\psi$ is convex even-symmetric , non-negative and non-decreasing on
$[0,+\infty)$, and $\psi(0)=0$. (ii) $\psi$ is twice differentiable on
$\mathbb{R}\setminus \{0\}$. (iii) $\psi$ is continuous on $\mathbb{R}$, it is not
necessarily smooth at zero and admits a positive right derivative at zero $\psi^{'}_+(0) >
0$.  Then, the proximity operator $\prox_{\delta\psi}(\beta) = \hat{\va}(\beta)$ has
exactly one continuous solution decoupled in each coordinate $\beta_i$ :
\begin{equation}
  \label{eq:10}
  \hat{\va}_i(\beta_i) =
  \begin{cases}
    0 & \text{if } \abs{\beta_i} \le \delta\psi^{'}_+(0)\\
    \beta_i-\delta\psi^{'}(\hat{\va}_i) & \text{if } \abs{\beta_i} > \delta\psi^{'}_+(0)
  \end{cases}
\end{equation}
\end{theorem}
See \cite{Fadili2006a}. Among the most popular penalty functions $\psi$ satisfying the
above requirements, we have $\psi(\va_i) = \abs{\va_i}$, in which case the associated
proximity operator is soft-thresholding. Therefore, \eqref{eq:5} is essentially an
iterative thresholding algorithm with a positivity constraint.

\vspace*{-0.4cm}
\section{Results}
\label{sec:results}
\vspace*{-0.4cm}

The performance of our approach has been assessed on several datasets of biological
images: a neuron phantom and a cell. Our algorithm was compared to RL with total variation
regularization (RL-TV \cite{Dey2004}), RL with multi-resolution support wavelet
regularization (RL-MRS \cite{Starck2006}), the naive proximal method that would assume the
noise to be Gaussian (NaiveGauss \cite{Vonesch2007}), and the approach of \cite{ChauxSPIE}
(AnsGauss). For all results presented, each algorithm was run with 200 iterations, enough
to reach convergence. Simulations were carried out with an approximated but realistic PSF
\cite{Zhang2006} whose parameters are obtained from a real confocal microscope. As usual,
the choice of $\lambda$ is crucial to balance between regularization and
deconvolution. For all the situations below, $\lambda$ was adjusted in order to reach the
best compromise.

\begin{figure}[ht]
  \centering
  \scriptsize{
  \begin{tabular}{@{ }c@{ }c@{ }c@{ }}
    \includegraphics[width=0.3\linewidth]{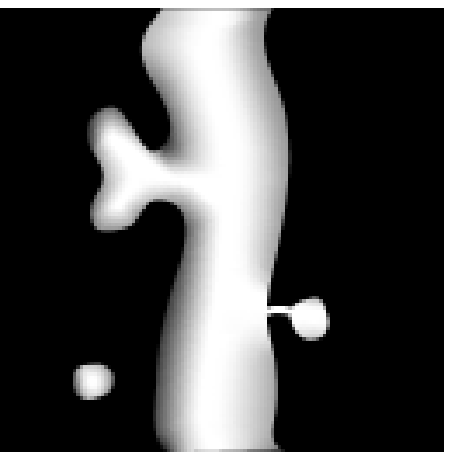} &
    \includegraphics[width=0.3\linewidth]{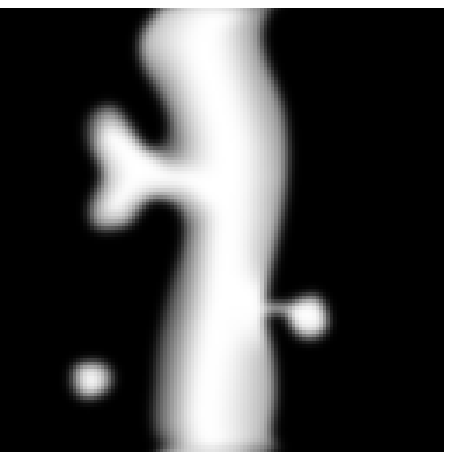} &
    \includegraphics[width=0.3\linewidth]{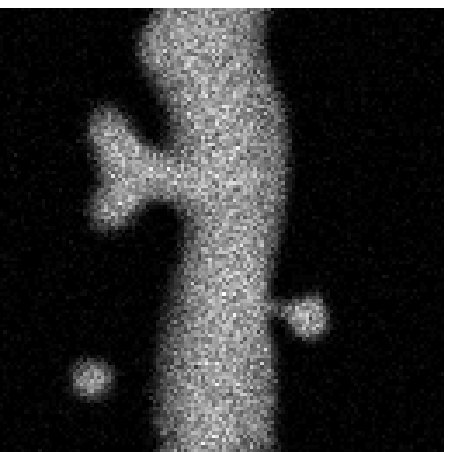} \\
    (a) & (b) & (c)	\\
    \includegraphics[width=0.3\linewidth]{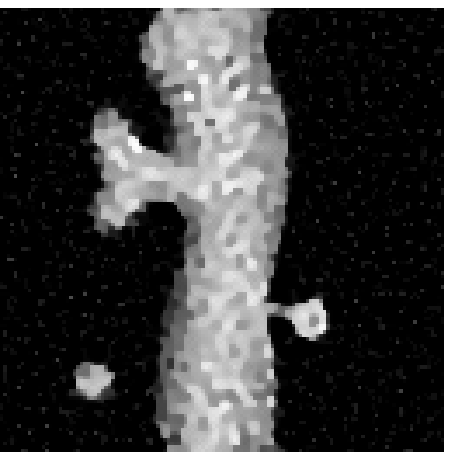} &
    \includegraphics[width=0.3\linewidth]{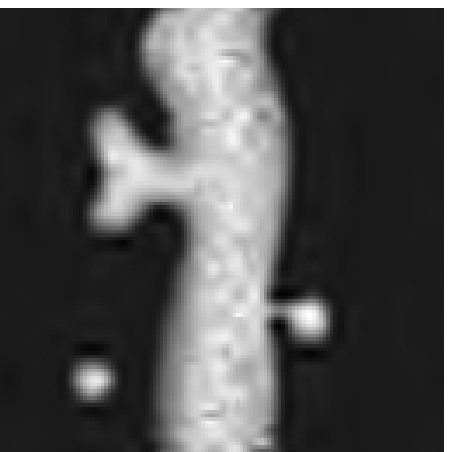} &
    \includegraphics[width=0.3\linewidth]{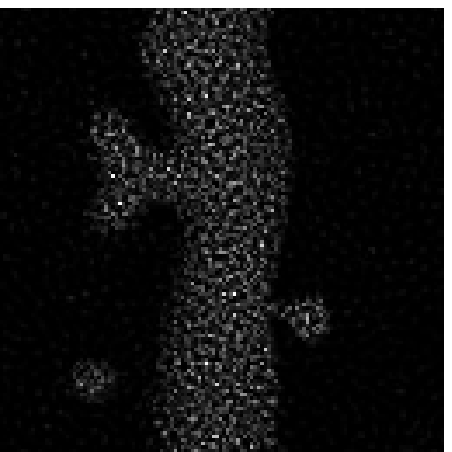}\\
    (d) & (e) & (f) \\
    \includegraphics[width=0.3\linewidth]{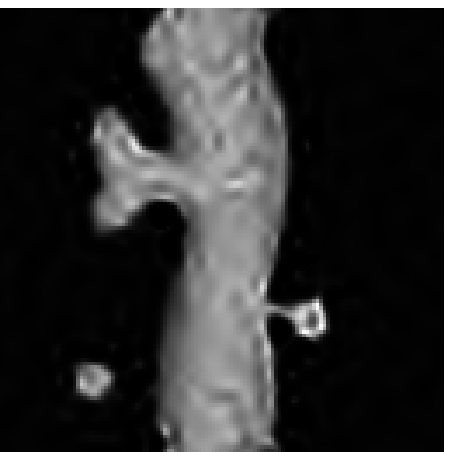} &
    \includegraphics[width=0.3\linewidth]{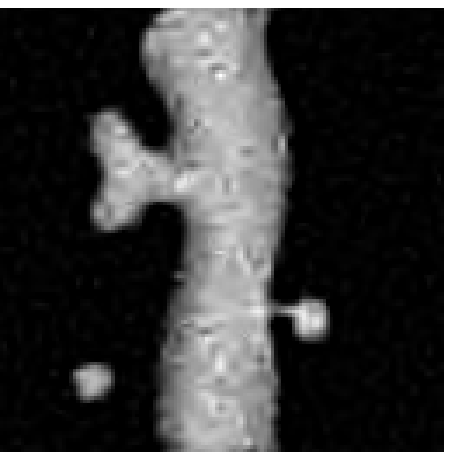}&  \\
    (g) & (h) &     \\
  \end{tabular}}
  \caption{\footnotesize{Deconvolution of a simulated neuron (Intensity $\leq$ 30). (a) Original, (b)
    Blurred, (c) Blurred\&noisy, (d) RL-TV, (e) NaiveGauss, (f) AnsGauss, (g) RL-MRS,
    (h) Our Algorithm.}}
  \label{fig:neuron}
\end{figure}

In Fig.\ref{fig:neuron}(a), a phantom of a neuron with a mushroom-shaped spine is
depicted.  The maximum intensity is 30. Its blurred and blurred+noisy versions are in (b)
and (c). With this neuron, and for NaiveGauss, AnsGauss and our approach, the dictionary
$\Phi$ contained the wavelet orthogonal basis. The deconvolution results are shown in
Fig.\ref{fig:neuron}(d)-(h). As expected the worst results are for the AnsGauss 
version, as it does not take care of the non-linearity of the Anscombe VST.
RL-TV shows rather good results but the background is full of artifacts. Our
approach provides a visually pleasant deconvolution result on this example. It efficiently
restores the spine, although the background is not fully cleaned.  RL-MRS also exhibits
good deconvolution results. Qualitative visual results are confirmed by quantitative
measures of the quality of deconvolution, where we used both the $\ell_1$-error (adapted
to Poisson noise), and the traditional MSE criteria. The $\ell_1$-errors for this image
are shown by Tab.~\ref{tab:intens-neuron} (similar results were obtained for the MSE). In
general, our approach performs well. It is competitive compared to RL-MRS which is
designed to directly handle Poisson noise. At low intensity regimes, our approach and
RL-MRS are the two algorithms that give the best results. At high intensity, RL-TV
performs very well, although RL-MRS, NaiveGauss and our approach are very close to it. NaiveGauss
performs poorly at low intensity as it does not correspond to a degradation model with Poisson
noise. AnsGauss gives the worst results probably because it does not handle the
non-linearity of the degradation model (\ref{eq:3}) after the VST.  To assess the
computational burden of the compared algorithms, Tab.~\ref{tab:time} summarizes the
execution times with an Intel PC Core 2 Duo 2GHz, 2Gb RAM. Except RL-MRS which is written
in C++, all other algorithms were implemented in MATLAB.

\begin{figure}[ht]
  \centering
  \scriptsize{
  \begin{tabular}{@{ }c@{ }c@{ }c@{ }}
    \includegraphics[width=0.3\linewidth]{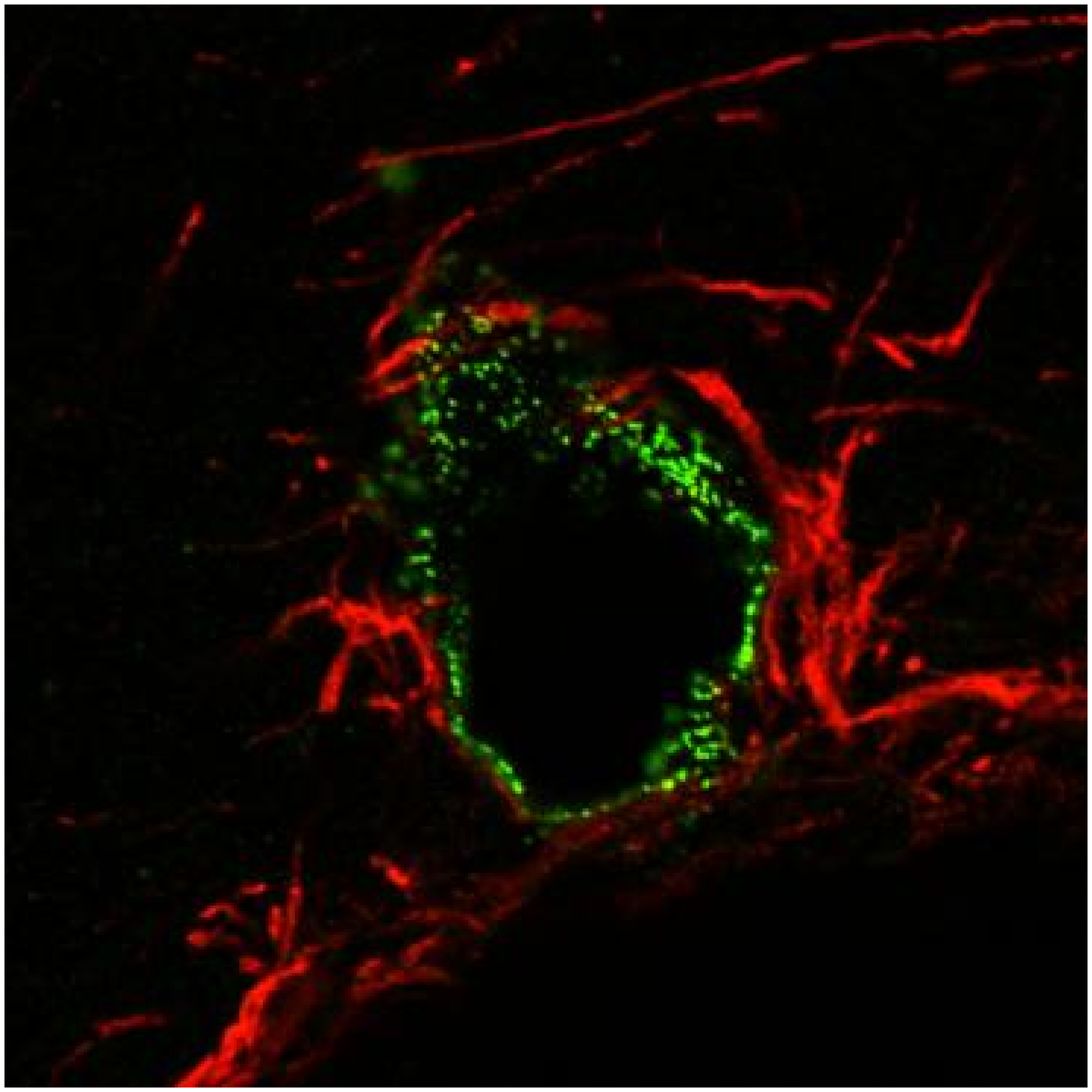} &
    \includegraphics[width=0.3\linewidth]{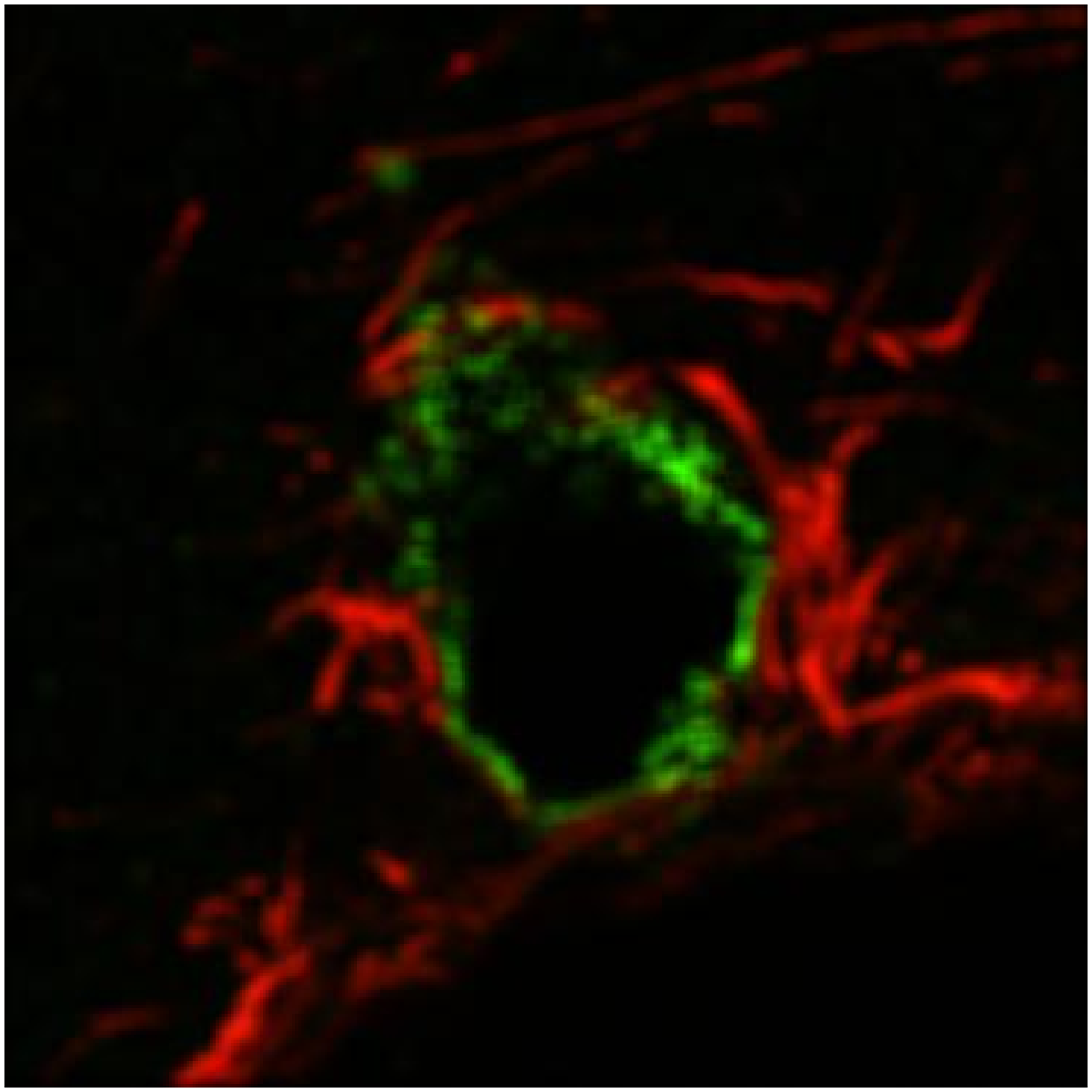} &
    \includegraphics[width=0.3\linewidth]{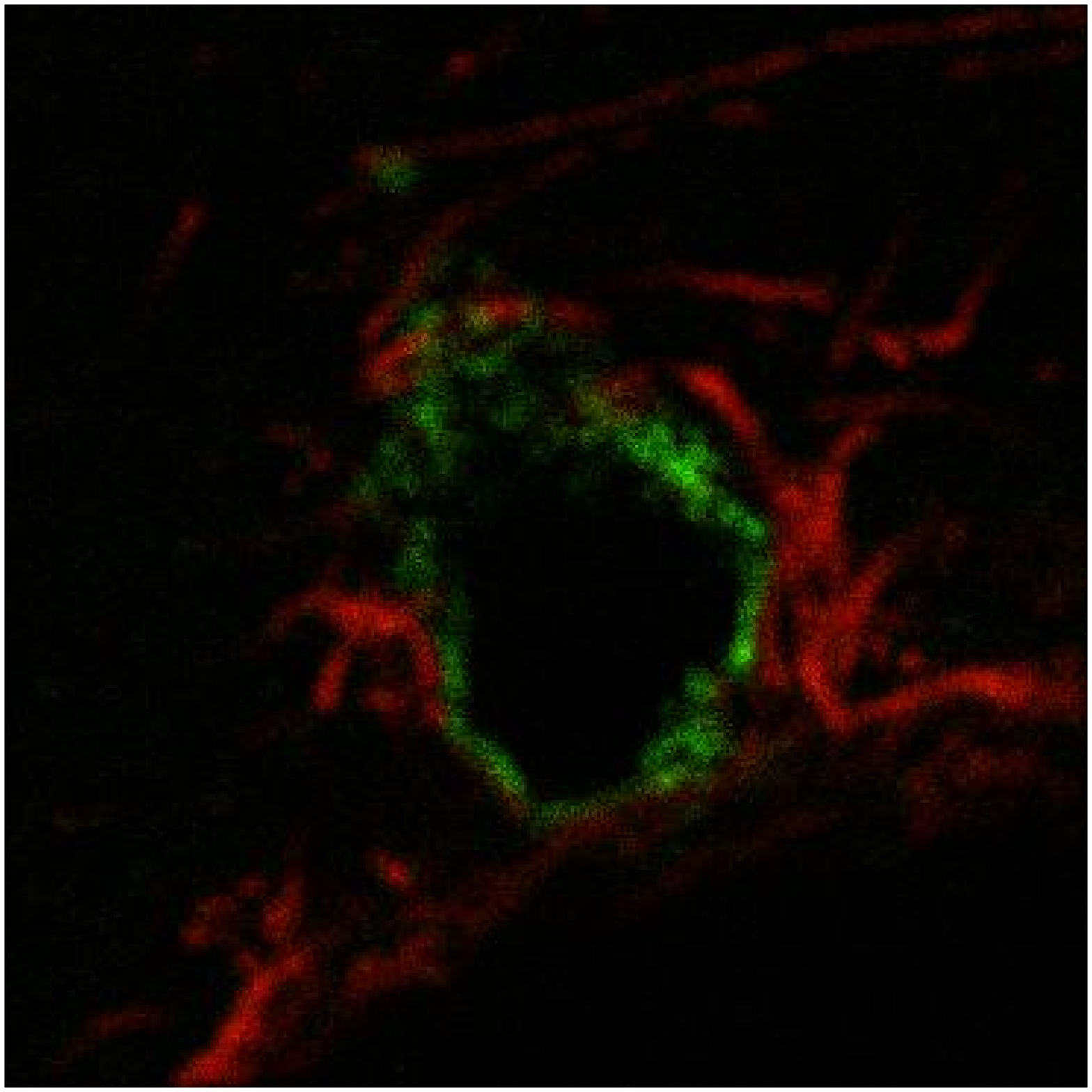} \\
    (a) & (b) & (c)	\\
    \includegraphics[width=0.3\linewidth]{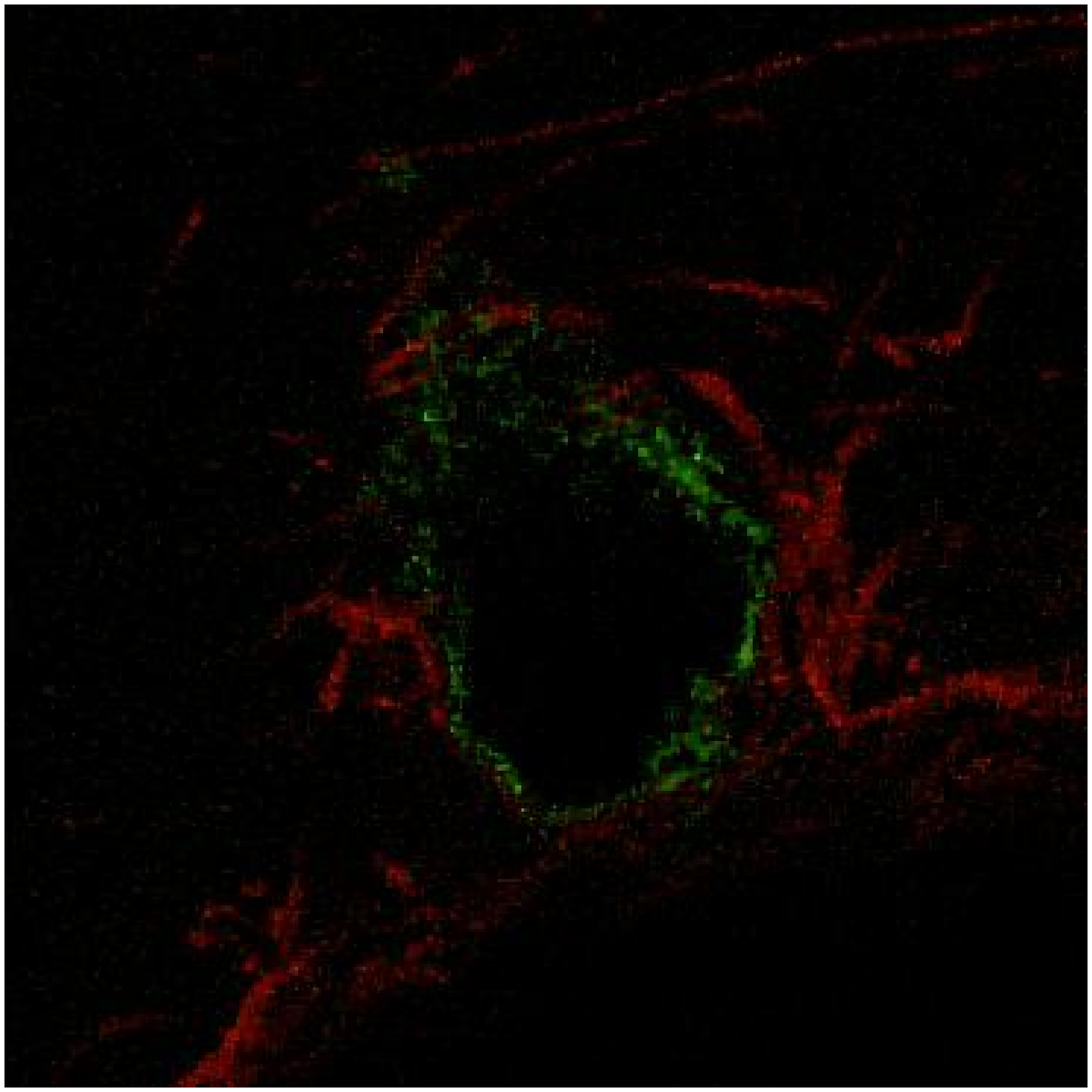} &
    \includegraphics[width=0.3\linewidth]{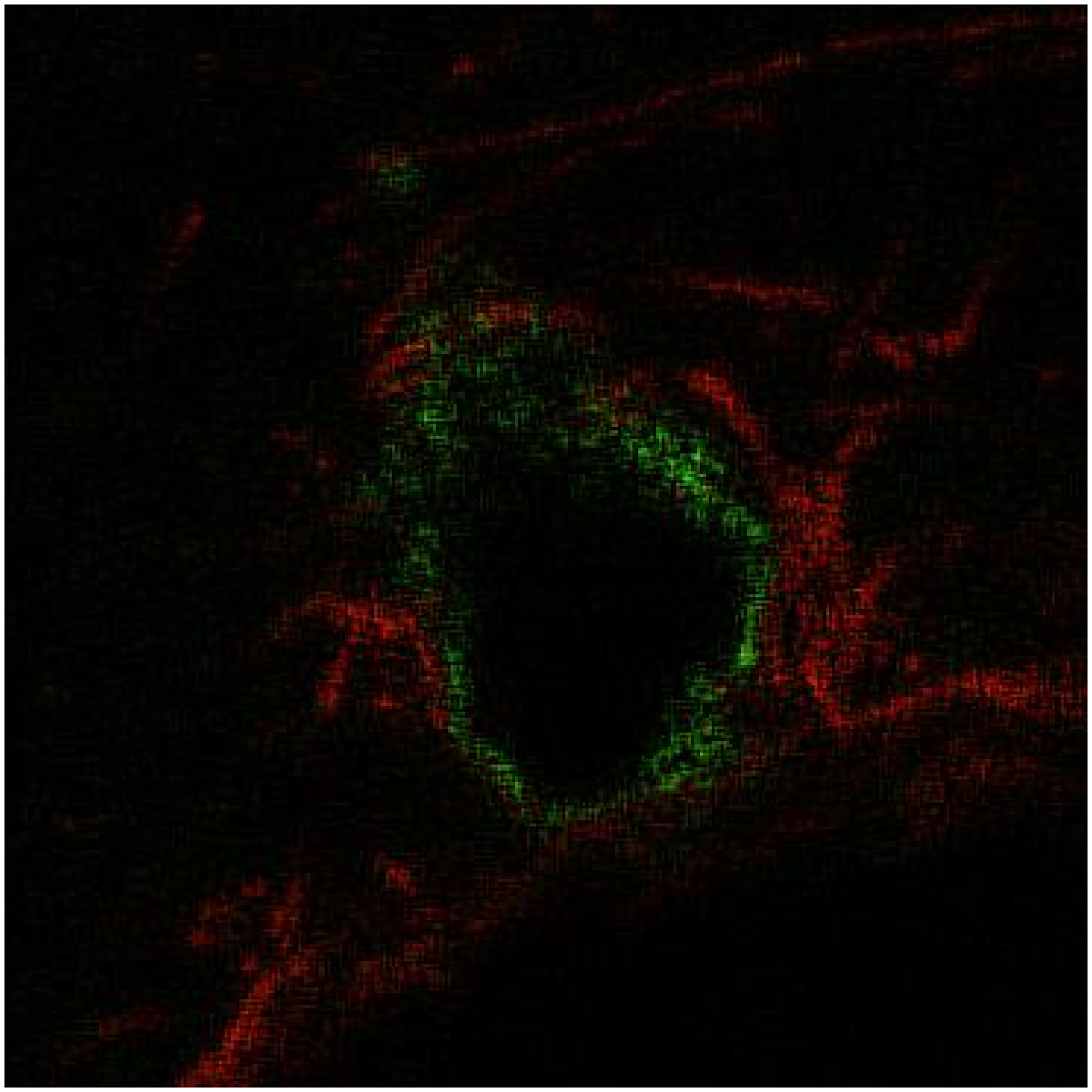} &
    \includegraphics[width=0.3\linewidth]{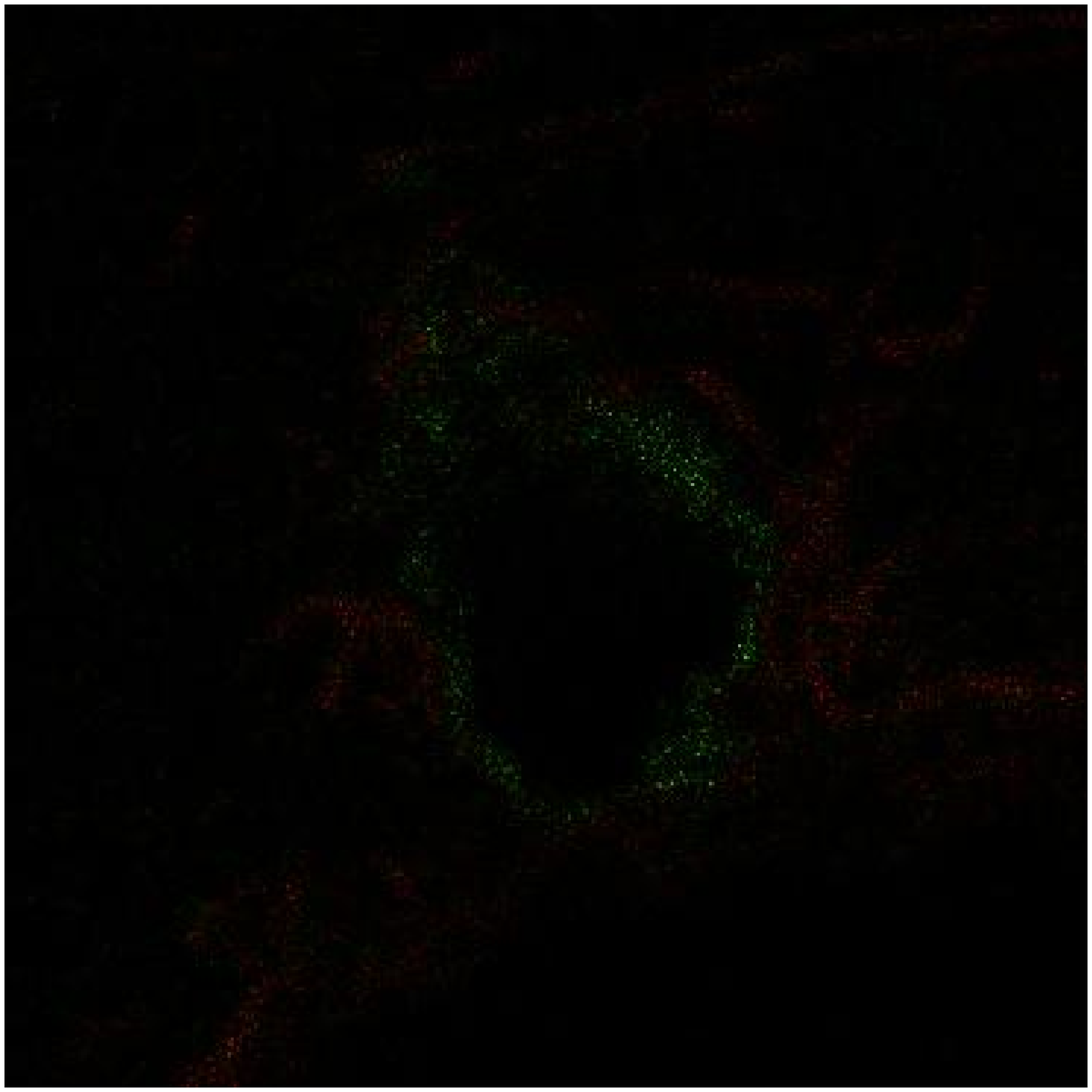}\\
    (d) & (e) & (f) \\
    \includegraphics[width=0.3\linewidth]{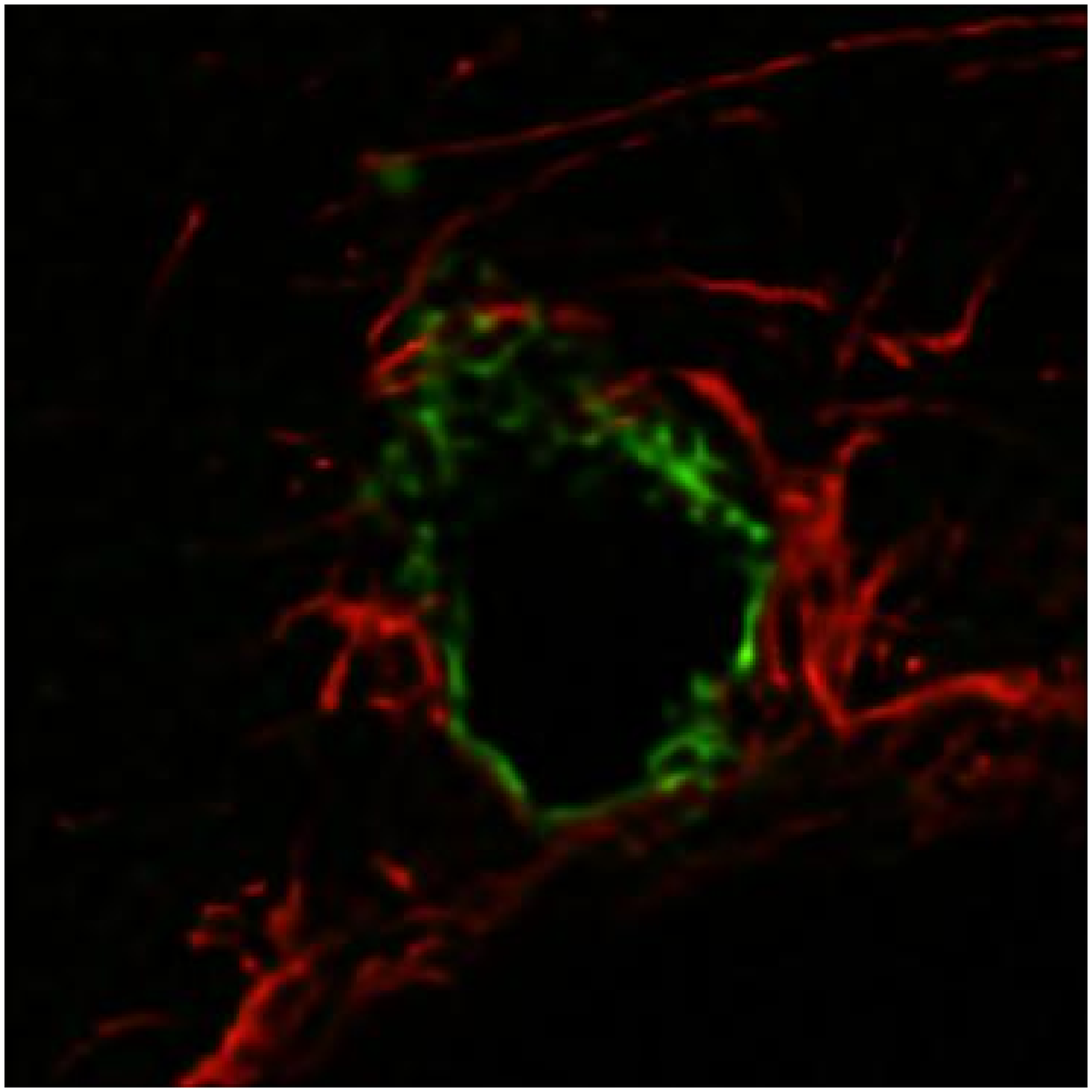} &
    \includegraphics[width=0.3\linewidth]{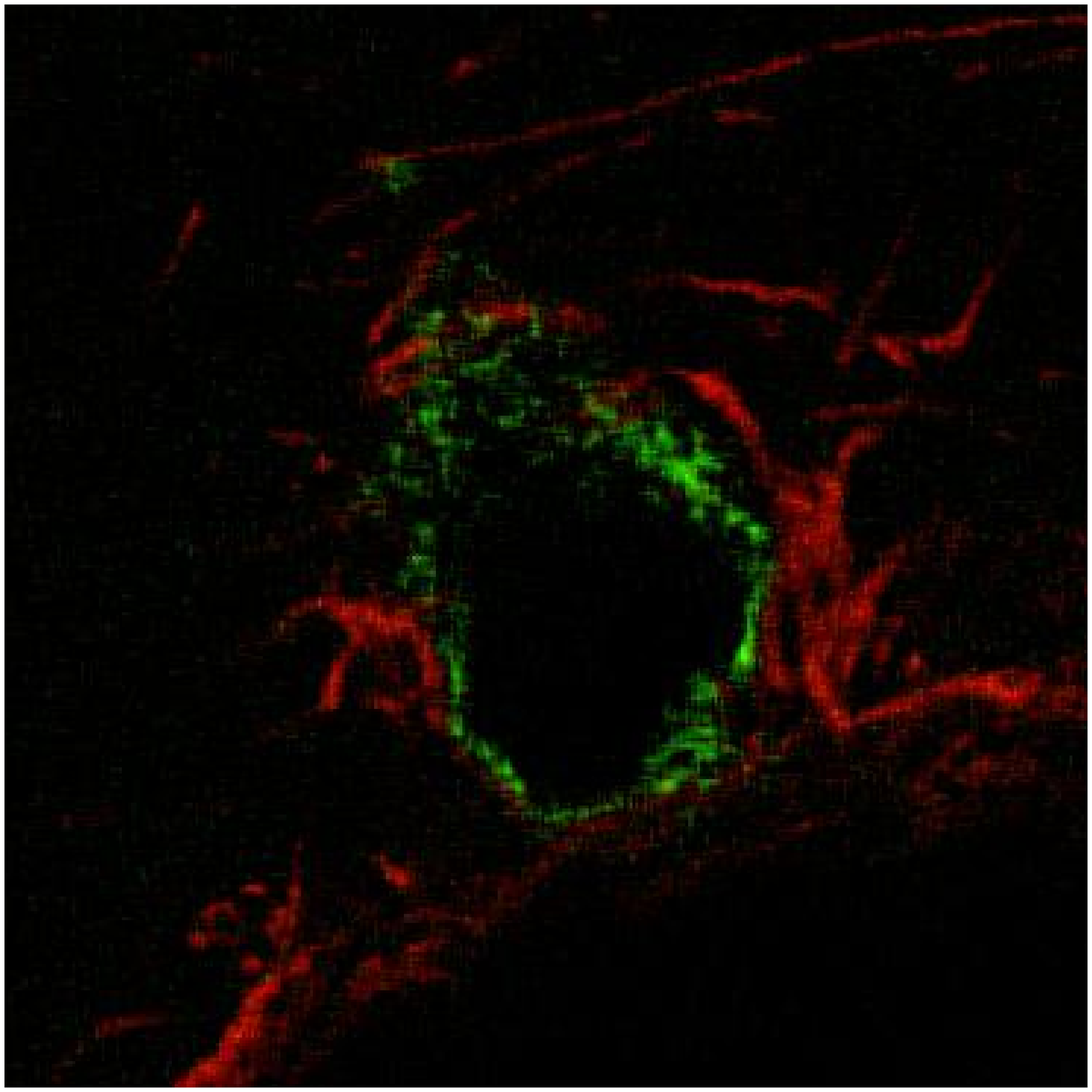}&  \\
    (g) & (h) &     \\
  \end{tabular}}
  \caption{\footnotesize{Deconvolution of the cell image (Intensity $\leq$ 30). (a) Original, (b)
    Blurred, (c) Blurred\&noisy, (d) RL-TV, (e) NaiveGauss, (f) AnsGauss, (g) RL-MRS,
    (h) Our Algorithm.}}
  \label{fig:cell}
\end{figure}

\begin{figure}[h]
  \centering
  \includegraphics[width=0.7\linewidth]{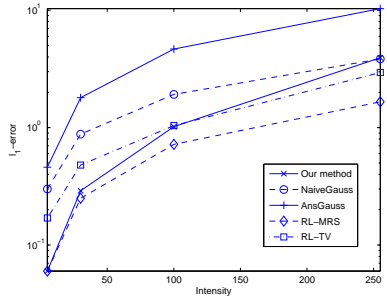}
  \caption{\footnotesize{Mean $\ell_1$-error of all algorithms as a function of the intensity level for the
    deconvolution of the cell}}
  \label{fig:graph-cell}
\end{figure}

The same experiment as above was carried out with a microscopy image of the endothelial
cell of the blood microvessel walls; see Fig.~\ref{fig:cell}.  For the NaiveGauss, AnsGauss
and our approach, the dictionary $\Phi$ contained the wavelet orthogonal basis and the
curvelet tight frame. The AnsGauss and the NaiveGauss results are spoiled by artifacts and
suffer from a loss of photometry.  RL-TV result shows a good restoration of small isolated
details but with a dominating staircase-like artifacts. RL-MRS and our approach give very
similar results although an extra-effort could be made to better restore tiny details. The
quantitative measures depicted in Fig.~\ref{fig:graph-cell} confirm this qualitative
discussion.

\begin{figure}[ht]
  \centering
  \footnotesize{
  \begin{tabular}{@{ }c@{ }c@{}}
    \includegraphics[width=0.45\linewidth]{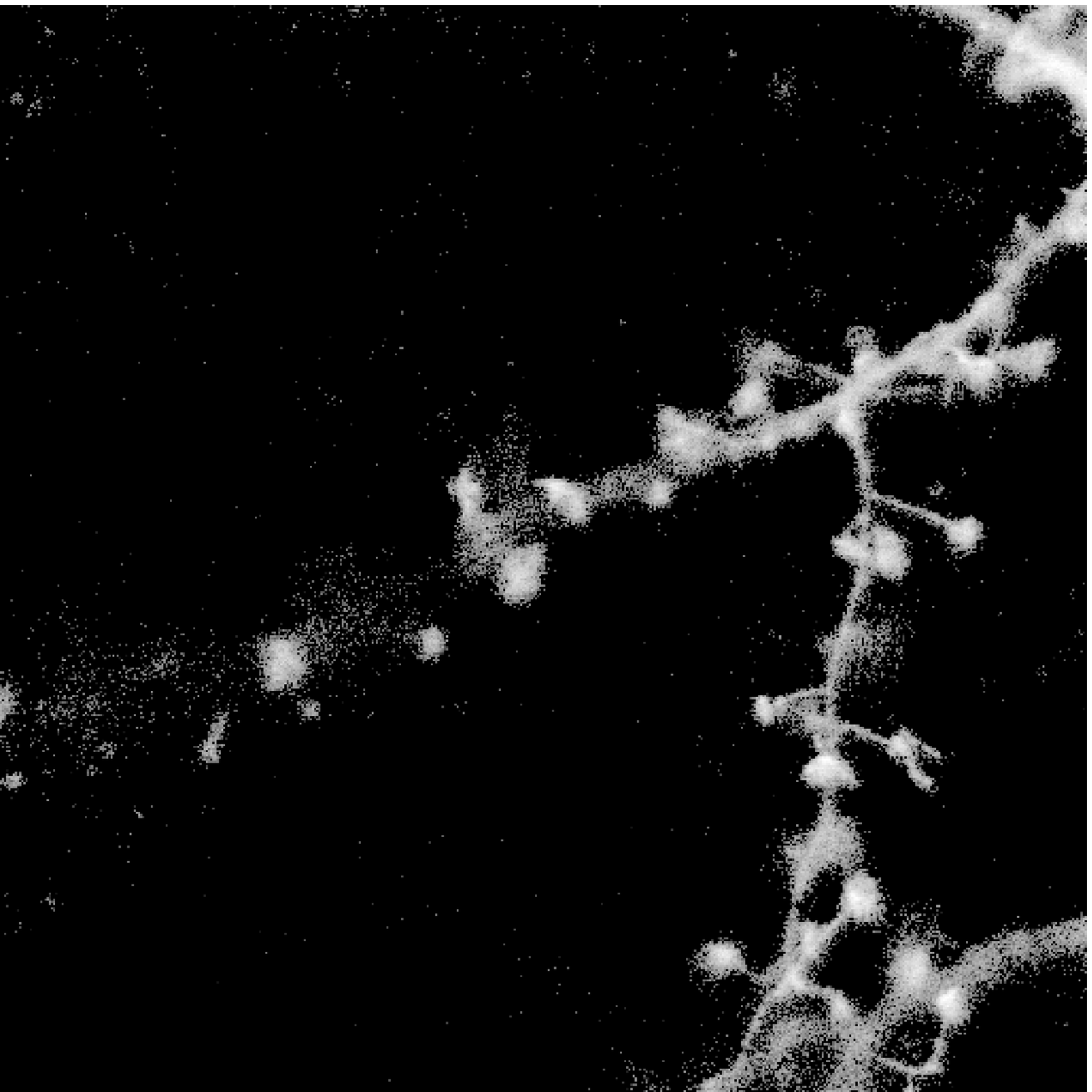} &
    \includegraphics[width=0.45\linewidth]{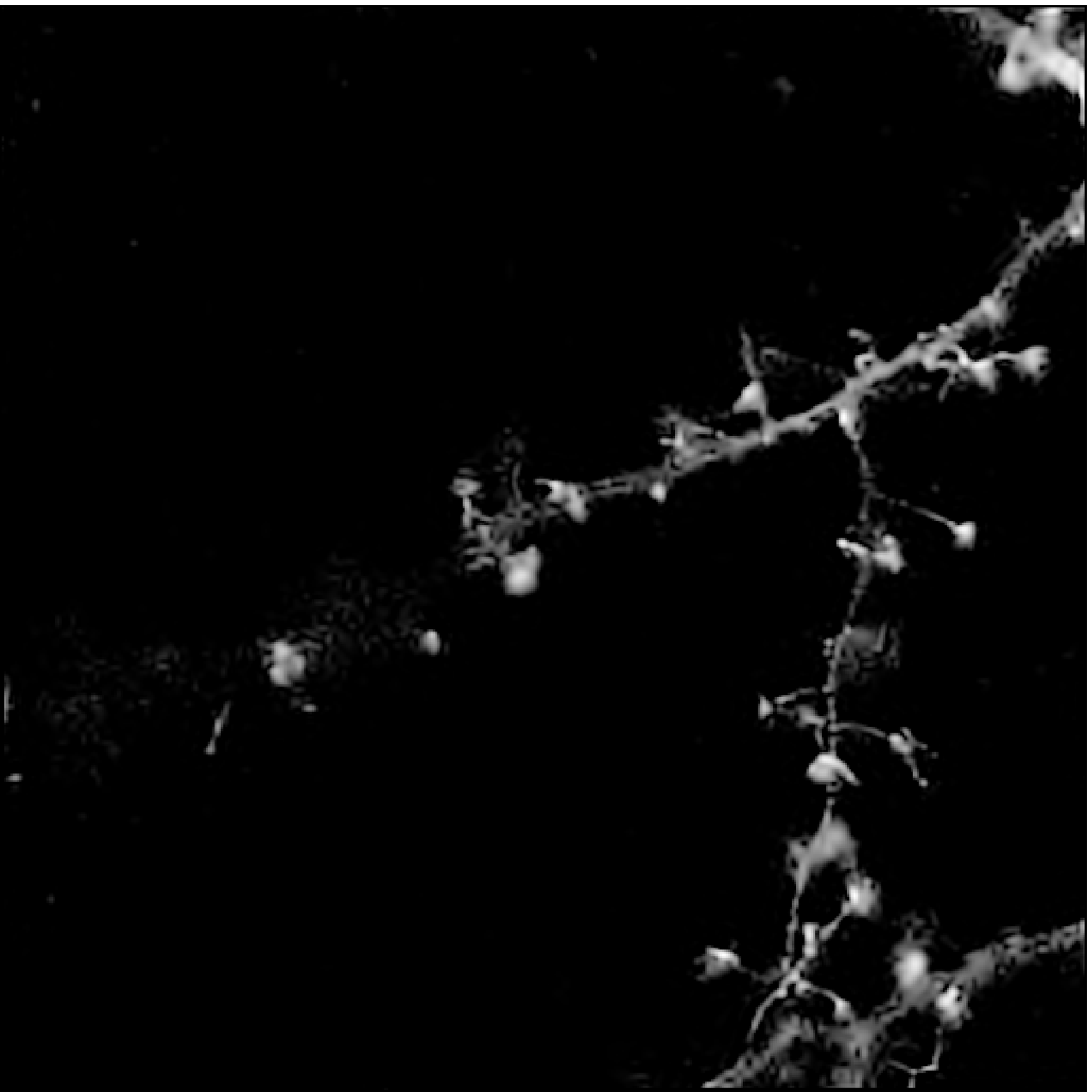} \\
    (a) & (b) \\
  \end{tabular}}
  \caption{\footnotesize{Deconvolution of a real neuron. (a) Original noisy, (b) Restored with our algorithm}}
  \label{fig:real-neuron}
\end{figure}

Finally, we applied our algorithm on a real confocal microscopy image of neurons.
Fig.~\ref{fig:real-neuron}(a) depicts the observed image\footnote{Courtesy of the GIP
  Cyc\'eron, Caen France.}  using the GFP fluorescent
protein. Fig.~\ref{fig:real-neuron}(b) shows the restored image using our algorithm with
the orthogonal wavelets. The images are shown in log-scale for visual purposes. We can
notice that the background has been cleaned and some structures have reappeared. The
spines are well restored and part of the dendritic tree is reconstructed, however some
information can be lost (see tiny holes). This can be improved using more relevant
transforms.

\vspace*{-0.2cm}

\begin{table}[ht]
  \centering
  \footnotesize{
  \begin{tabular}{|c|c|c|c|c|}
    \hline  & \multicolumn{4}{|c|}{Intensity regime}\\
    \hline Method		        & $\leq 5$ & $\leq 30$ & $\leq 100$ & $\leq 255$ \\
    \hline Our method			&  0.21    &  0.76     & 2.39       & 4.79       \\
    \hline NaiveGauss			&  0.41    &  0.89     & 2.31       & 5.15       \\
    \hline AnsGauss			&  2.07    &  7.56     & 21.25      & 50.41      \\
    \hline RL-MRS                       &  0.21    &  0.96     & 2.2        & 4.51       \\
    \hline RL-TV                        &  0.73    &  1.47     & 2.72       & 4.28       \\
    \hline
  \end{tabular}}
  \caption{\footnotesize{Mean $\ell_1$-error of all algorithms as a function of the intensity level for the
    deconvolution of the neuron phantom.}}
  \label{tab:intens-neuron}
\end{table}

\vspace*{-0.5cm}
\begin{table}[ht]
  \centering
  \footnotesize{
  \begin{tabular}{|c|c|}
    \hline Method		        &  Time (in s)\\
    \hline Our method			&  2.7        \\
    \hline NaiveGauss			&  1.7        \\
    \hline AnsGauss			&  1.7        \\
    \hline RL-MRS                       &  15         \\
    \hline RL-TV                        &  2.5        \\
    \hline
  \end{tabular}}
  \caption{\footnotesize{Execution time for the simulated neuron using the orthogonal wavelet transform}}
  \label{tab:time}
\end{table}


\vspace*{-0.4cm}
\section{Conclusion}
\label{sec:conclusion}
\vspace*{-0.4cm}

In this paper, we presented a sparsity-based fast iterative thresholding deconvolution
algorithm that take accounts of the presence of Poisson noise. Competitive results on
confocal microscopy images with state-of-the-art algorithms are shown. The combination of
several transforms leads to some advantages, as we can easily
adapt the dictionary to the kind of image to restore. The parameter $\lambda$ can be
tricky to find, and we are developing a method helping to solve this issue.

\vspace*{-0.5cm}
\footnotesize
\bibliographystyle{IEEEbib}
\bibliography{paperbib}

\end{document}

%% file: prelim.tex
\newtheorem{theorem}{Theorem}

\newcommand{\va}{{{\alpha}}}

\newcommand{\I}{\mathbf{I}}
\newcommand{\vx}{{x}}

\newcommand{\R}{\mathbb{R}}

\newcommand{\T}{\mathbf{T}}

\newcommand{\be}{\begin{eqnarray}}
\newcommand{\ee}{\end{eqnarray}}

\newcommand{\M}{\mathcal{M}}
\renewcommand{\P}{\mathsf{P}}

\newcommand{\C}{\mathcal{C}}

\newcommand{\norm}[1]{\left\|#1\right\|}
\newcommand{\abs}[1]{\left|#1\right|}

\newcommand{\parenth}[1]{\left({#1}\right)}

%% file: papier.bbl
\begin{thebibliography}{10}

\bibitem{Dey2004}
N.~Dey et~al.,
\newblock ``A deconvolution method for confocal microscopy with total variation
  regularization,''
\newblock in {\em IEEE ISBI}, 2004.

\bibitem{Sarder2006}
P.~Sarder and A.~Nehorai,
\newblock ``{Deconvolution Method for {3-D} Fluorescence Microscopy Images},''
\newblock {\em IEEE Sig. Pro.}, vol. 23, pp. 32--45, 2006.

\bibitem{Monvel2001}
J.~Boutet de~Monvel~et al,
\newblock ``{Image Restoration for Confocal Microscopy: Improving the Limits of
  Deconvolution, with Application of the Visualization of the Mammalian Hearing
  Organ},''
\newblock {\em Biophysical Journal}, vol. 80, pp. 2455--2470, 2001.

\bibitem{Vonesch2007}
C.~Vonesch and M.~Unser,
\newblock ``A fast iterative thresholding algorithm for wavelet-regularized
  deconvolution,''
\newblock {\em IEEE ISBI}, 2007.

\bibitem{Daubechies2004}
I.~Daubechies, M.~Defrise, and C.~De Mol,
\newblock ``An iterative thresholding algorithm for linear inverse problems
  with a sparsity constraints,''
\newblock {\em CPAM}, vol. 112, pp. 1413--1541, 2004.

\bibitem{Combettes2005}
P.~L. Combettes and V.~R. Wajs,
\newblock ``Signal recovery by proximal forward-backward splitting,''
\newblock {\em SIAM MMS}, vol. 4, no. 4, pp. 1168--1200, 2005.

\bibitem{Teschke2007}
G.~Teschke,
\newblock ``Multi-frame representations in linear inverse problems with mixed
  multi-constraints,''
\newblock {\em ACHA}, vol. 22, no. 1, pp. 43--60, 2007.

\bibitem{Combettes2007b}
C.~Chaux, P.~L. Combettes, J.-C. Pesquet, and V.~R. Wajs,
\newblock ``A variational formulation for frame-based inverse problems,''
\newblock {\em Inv. Prob.}, vol. 23, pp. 1495--1518, 2007.

\bibitem{Fadili2006a}
M.~J. Fadili and J.-L. Starck,
\newblock ``Sparse representation-based image deconvolution by iterative
  thresholding,''
\newblock in {\em ADA IV}, France, 2006, Elsevier.

\bibitem{ChauxSPIE}
C.~Chaux, L.~Blanc-F\'eraud, and J.~Zerubia,
\newblock ``Wavelet-based restoration methods: application to 3d confocal
  microscopy images,''
\newblock in {\em SPIE Wavelets XII}, 2007.

\bibitem{Dupe2008}
F-X Dup\'e, M.J. Fadili, and J.-L. Starck,
\newblock ``A proximal iteration for deconvolving poisson noisy images using
  sparse representations,''
\newblock {\em IEEE Trans. Im. Proc.}, 2008,
\newblock submitted.

\bibitem{Starck2006}
J.-L. Starck and F.~Murtagh,
\newblock {\em Astronomical Image and Data Analysis},
\newblock Springer, 2006.

\bibitem{Zhang2006}
B.~Zhang, J.~Zerubia, and J.-C. Olivo-Marin,
\newblock ``{Gaussian approximations of fluorescence microscope PSF models},''
\newblock {\em OSA}, 2006.

\end{thebibliography}
